\documentstyle[12pt]{article}
\newlength{\extraspace}
\newlength{\extraspaces}
\newcommand{\be}{\begin{equation}
\addtolength{\abovedisplayskip}{\extraspaces}
\addtolength{\belowdisplayskip}{\extraspaces}
\addtolength{\abovedisplayshortskip}{\extraspace}
\addtolength{\belowdisplayshortskip}{\extraspace}}
\newcommand{\ee}{\end{equation}}

\newcommand{\ba}{\begin{eqnarray}
\addtolength{\abovedisplayskip}{\extraspaces}
\addtolength{\belowdisplayskip}{\extraspaces}
\addtolength{\abovedisplayshortskip}{\extraspace}
\addtolength{\belowdisplayshortskip}{\extraspace}}
\newcommand{\ea}{\end{eqnarray}}

\newcommand{\newsection}[1]{
\vspace{15mm}
\pagebreak[3]
\addtocounter{section}{1}
\setcounter{equation}{0}
\setcounter{subsection}{0}
\setcounter{footnote}{0}
\begin{flushleft}
{\large\bf \thesection. #1}
\end{flushleft}
\nopagebreak
\medskip
\nopagebreak}

\newcommand{\Tr}{{\rm Tr}}
\newcommand{\Dmrns}{{\cal D}_{\mu\rho,\nu\sigma}}

\begin{document}

\addtolength{\baselineskip}{.8mm}

{\thispagestyle{empty}
%

\begin{center}
\vspace*{1.0cm}
{\large\bf Gauge--invariant field--strength correlators in pure}\\
{\large\bf Yang--Mills and full QCD at finite temperature
 \footnote{Partially supported by MIUR (Italian Ministry of the University 
 and of Scientific and Technological Research) and by the INTAS contract
 00--0110.} }\\
\vspace*{1.0cm}
{\large M. D'Elia}\\
\vspace*{0.5cm}{\normalsize
{Dipartimento di Fisica,\\
Universit\`a di Genova,\\
and INFN, Sezione di Genova,\\
I--16146 Genova, Italy.}}\\
\vspace*{1.0cm}
{\large A. Di Giacomo, E. Meggiolaro}\\
\vspace*{0.5cm}{\normalsize
{Dipartimento di Fisica, \\
Universit\`a di Pisa, \\
and INFN, Sezione di Pisa,\\
I--56127 Pisa, Italy.}}\\
\vspace*{1.0cm}{\large \bf Abstract}
\end{center}
\noindent
We study by numerical simulations on a lattice the behaviour of
the gauge--invariant two--point correlation functions of the gauge--field
strengths across the deconfinement phase transition, both for the pure--gauge
$SU(3)$ theory and for full QCD with two flavours.
{\it Quenched} data agree within errors with previous determinations, 
but have much higher statistics.
A best--fit analysis of the data has been performed, both for the
{\it quenched} and the full--QCD case, showing that the electric gluon
condensate drops to zero at the deconfining phase transition.
\\
\vspace{1.0cm}
\noindent
(PACS codes: 12.38.Gc, 11.10.Wx)
}
\vfill\eject

\newsection{Introduction}

\noindent
Gauge--invariant correlation functions of the field strenghts in the QCD
vacuum play an important role in high--energy phenomenology and in stochastic 
models
of QCD, both at zero temperature
\cite{Dosch87,Dosch88,Simonov89} and non--zero temperature
\cite{Simonov1,Simonov2,Simonov3}.
For a recent review see Ref. \cite{DDSS02}.
Some years ago, a determination of such correlators at finite temperature
was done on a $16^3 \times 4$ lattice, for the pure--gauge $SU(3)$ theory,
in a range of distances from $0.4$ to 1 fm approximately
\cite{npb97,DiGiacomo96}.
The technique used to make the computation feasible is a local cooling of
the configurations: this procedure freezes local fluctuations, leaving
long--range correlations unchanged.
In this paper, prompted by the progresses of the stochastic--vacuum
approach to QCD~\cite{Simonov4}, we improve the determination of the
correlators at finite temperature $T \sim T_c$ for pure gauge $SU(3)$,
by use of a larger lattice ($32^3 \times 6$) and bigger statistics.
We also compute the correlators at $T=0$ and at $T \sim T_c$ in full
QCD with 2 staggered dynamical quarks.
In Sect. 2 we recall the notation. The numerical results are presented
in Sect. 3, while Sect. 4 contains an analysis of the data and a discussion.

\newsection{Notation}

\noindent
To simulate the system at finite temperature, a lattice is used of spatial
extent $N_\sigma \gg N_\tau$, $N_\tau$ being the temporal extent,
with periodic boundary conditions for gluons, and antiperiodic
boundary conditions for fermions in the temporal direction.
The temperature $T$ corresponding to a given value of $\beta = 6/g^2$ is
given by
\be
N_\tau \cdot a = {1 \over T} ~,
\label{ntau}
\ee
where $a$ is the lattice spacing.
In the {\it quenched} case $a$ only depends on the coupling
$\beta$ and, from renormalization group arguments,
\be
a(\beta) = {1\over\Lambda_L} f(\beta) ~,
\label{abeta}
\ee
where $f(\beta)$ is the so--called {\it scaling function} and $\Lambda_L$
is the scale parameter of QCD in the lattice regularization scheme.
At large enough $\beta$, $f(\beta)$ is given by the usual two--loop
expression:
\be
f(\beta) = \left({8\over33}\,\pi^2\beta\right)
^{ 51/121 } \exp\left(-{4\over33}\pi^2\beta\right)
\left[1+{\cal O}(1/\beta)\right] ~,
\label{2-loop}
\ee
for gauge group $SU(3)$ and in the absence of quarks.
The expression (\ref{2-loop}) can also be used in a small enough interval of
$\beta$'s lower than the asymptotic scaling region, and then $\Lambda_L$ is an
effective scale depending on the position of the interval considered.
For the range of values of $\beta$'s that we have considered (see below),
its value, extracted from the string tension
\cite{Michael88,Bali-Schilling93}, is about 4.9 MeV.

The gauge--invariant two--point correlators of the field strengths in the
QCD vacuum are defined as~\cite{Dosch87,Dosch88,Simonov89}
\be
\Dmrns(x) = g^2 \langle
\Tr [ G_{\mu\rho}(0) S(0,x) G_{\nu\sigma}(x) S^\dagger(0,x) ]
\rangle ~,
\label{defcorr}
\ee
where $G_{\mu\rho} = T^aG^a_{\mu\rho}$ is the field--strength tensor,
$T^a$ are the generators of the algebra of the gauge group
in the fundamental representation, and
\be
S(0,x) = {\rm P}\exp\left(i g \int^1_0dt\,x^\mu A_\mu(xt)\right)
\label{string}
\ee
is the Schwinger parallel transport from $0$ to $x$ along a straight--line
path;\footnote{Recently a strong dependence of the correlators on the shape
of the path in the Schwinger string has been found numerically \cite{plb2002}.
However, the stochastic--vacuum models naturally select the straight--line
path. For that reason, we limit our present analysis to the straight--line
Schwinger string.}
$A_\mu \equiv T^aA^a_\mu$.
At zero temperature, that is on a symmetric lattice $N_\sigma = N_\tau$,
the correlators are expressed in terms of two independent invariant functions
of $x^2$, known as ${\cal D} (x^2)$ and ${\cal D}_1 (x^2)$
~\cite{Dosch87,Dosch88,Simonov89}:
\ba
\lefteqn{
\Dmrns(x) = (\delta_{\mu\nu}\delta_{\rho\sigma} - \delta_{\mu\sigma}
\delta_{\rho\nu})
\left[ {\cal D}(x^2) + {\cal D}_1(x^2) \right] } \nonumber \\
& & + (x_\mu x_\nu \delta_{\rho\sigma} - x_\mu x_\sigma \delta_{\rho\nu}
+ x_\rho x_\sigma \delta_{\mu\nu} - x_\rho x_\nu \delta_{\mu\sigma})
{\partial{\cal D}_1(x^2) \over \partial x^2} ~.
\label{param}
\ea
At finite temperature ($N_\sigma \gg N_\tau$) the $O(4)$ space--time symmetry
is broken down to the spatial $O(3)$ symmetry and the bilocal correlators are
now expressed in terms of five independent functions
\cite{Simonov1,Simonov2,Simonov3}. Two of them are needed to
describe the electric--electric correlations:
\ba
\lefteqn{
g^2 \langle \Tr [ E_i (x) S(x,y) E_k (y) S^\dagger(x,y) ]
\rangle } \nonumber \\
& & = \delta_{ik} \left[ D^E + D_1^E + u_4^2 {\partial D_1^E \over
\partial u_4^2} \right] + u_i u_k {\partial D_1^E \over
\partial \vec{u}^2} ~,
\label{EE}
\ea
where $E_i = G_{i4}$ is the electric field operator and
$u_\mu = x_\mu - y_\mu$ [$\vec{u}^2 = (\vec{x} - \vec{y})^2$].

Two further functions are needed for the magnetic--magnetic correlations:
\ba
\lefteqn{
g^2 \langle \Tr [ B_i (x) S(x,y) B_k (y) S^\dagger(x,y) ]
\rangle } \nonumber \\
& & = \delta_{ik} \left[ D^B + D_1^B + \vec{u}^2 {\partial D_1^B \over
\partial \vec{u}^2} \right] - u_i u_k {\partial D_1^B \over
\partial \vec{u}^2} ~,
\label{BB}
\ea
where $B_k = {1 \over 2} \varepsilon_{ijk} G_{ij}$ is the magnetic field
operator.

Finally, one more function is necessary to describe the mixed
electric--magnetic correlations:
\be
g^2 \langle \Tr [ E_i (x) S(x,y) B_k (y) S^\dagger(x,y) ]
\rangle = -{1 \over 2} \varepsilon_{ikn} u_n
{\partial D_1^{BE} \over \partial u_4} ~.
\label{EB}
\ee
In Eqs. (\ref{EE}), (\ref{BB}) and (\ref{EB}), the five quantities $D^E$,
$D_1^E$, $D^B$, $D_1^B$ and $D_1^{BE}$ are all functions of $\vec{u}^2$,
due to rotational invariance, and of $u_4^2$, due to time--reversal invariance.

From the conclusions of Refs. \cite{Simonov1,Simonov2,Simonov3},
one expects that $D^E$ is related to the (temporal)
string tension and should have a drop just above the deconfinement critical
temperature $T_c$. In other words, $D^E$ is expected to be a kind of order
parameter for confinement; on the contrary, $D^E_1$ does not contribute to
the area law of the temporal Wilson loop.
Similarly, $D^B$ is related to the spatial string tension
\cite{Simonov1,Simonov2}, while $D^B_1$ does not contribute to the area
law of the spatial Wilson loop.

The above arguments hold both for the {\it quenched} and the {\it unquenched}
theory, with a suitable modification of Eq. (\ref{abeta}) and (\ref{2-loop}).
In particular for the full--QCD case, in order to fix the scale, we have
used the lattice spacing as determined in~\cite{pisa2000}.

\newpage

\newsection{Results}

\noindent
We have determined the following four quantities \cite{DiGiacomo92}
\ba
{\cal D}_\parallel^E (\vec{u}^2,0) &\equiv&
 {\cal D}^E (\vec{u}^2,0) + {\cal D}_1^E (\vec{u}^2,0) +
\vec{u}^2 {\partial{\cal D}_1^E \over \partial \vec{u}^2} (\vec{u}^2,0) ~;
\nonumber \\
{\cal D}_\perp^E (\vec{u}^2,0) &\equiv& {\cal D}^E (\vec{u}^2,0)
+ {\cal D}_1^E (\vec{u}^2,0) ~;
\label{corr-E}
\\
{\cal D}_\parallel^B (\vec{u}^2,0) &\equiv&
 {\cal D}^B (\vec{u}^2,0) + {\cal D}_1^B (\vec{u}^2,0)
+ \vec{u}^2 {\partial{\cal D}_1^B \over \partial \vec{u}^2} (\vec{u}^2,0) ~;
\nonumber \\
{\cal D}_\perp^B (\vec{u}^2,0) &\equiv& {\cal D}^B (\vec{u}^2,0) +
{\cal D}_1^B (\vec{u}^2,0) ~,
\label{corr-B}
\ea
by measuring appropriate linear superpositions of the correlators (\ref{EE})
and (\ref{BB}) at equal times ($u_4 = 0$).
Concerning the mixed electric--magnetic correlator of Eq. (\ref{EB}),
it vanishes both at zero temperature and at finite temperature, when computed
at equal times ($u_4 = 0$), as a consequence of the invariance of the theory
under time reversal.

We have chosen a $32^3 \times 6$ lattice (so that, in our notation,
$N_\tau = 6$) for the {\it quenched} case.
The critical temperature $T_c$ for such a lattice corresponds to
$\beta_c \simeq 5.89$~\cite{Boyd96}.
For full QCD we have used 2 species of staggered
fermions with bare mass $a \cdot m = 0.0125$ and a $32^3 \times 8$ lattice,
for which $T_c$ corresponds to a coupling $\beta_c \simeq 5.54$~\cite{gott93}.
The standard ``R--version'' of the HMC algorithm~\cite{gott87} has been used
in the full--QCD case.

For the {\it quenched} theory the behaviour of ${\cal D}_\parallel^E$ and
${\cal D}_\perp^E$ is shown in Figs. 1 and 2 respectively, at different values
of $T/T_c$ with the physical distance in the range from $\sim 0.25$ fm up to
$\sim 1.25$ fm.
A clear drop is observed for ${\cal D}_\parallel^E$ and
${\cal D}_\perp^E$ across the phase transition, as expected.
The analogous behaviour for ${\cal D}_\parallel^B$ and ${\cal D}_\perp^B$
is shown in Figs. 3 and 4. No dramatic change is visible across the transition
for the magnetic correlations.
These behaviours of the correlators were already known from Refs.
\cite{npb97,DiGiacomo96}, with which we agree within the errors.
In Figs. 1--4 the thick continuum line has been obtained using the parameters
of the best fit at $T=0$ obtained in Ref. \cite{npb97}, Eqs. (2.10) and (2.11).

In Figs. 5--8 we present the analogous data for the case with 
dynamical fermions. In full QCD previous results at $T = 0$ have
been reported only for the case of 4 flavours~\cite{plb97}. Therefore,
in order to compare with $T = 0$, we have performed a simulation 
at $\beta = 5.55$ on a $16^4$ lattice: the results are reported in the
same figures.
Also in this case it is apparent that ${\cal D}_\parallel^E$ and
${\cal D}_\perp^E$ stay almost constant at their zero--temperature value
up to the phase transition, where they undergo a sharp drop. 
Instead no dramatic change is visible for the magnetic correlators.


Our results, both for the {\it quenched} and the full--QCD case, are
in agreement with those already found in Refs. \cite{npb97,DiGiacomo96}
and can be summarized as follows:
\begin{itemize}
\item[(1)] In the confined phase ($T < T_c$), up to temperatures very near to
$T_c$, the correlators, both the electric--electric type (\ref{EE}) and
the magnetic--magnetic type (\ref{BB}), are nearly equal to the correlators at
zero temperature: in other words, $D^E \simeq D^B \simeq D$ and
$D_1^E \simeq D_1^B \simeq D_1$ for $T < T_c$.
\item[(2)] Immediately above $T_c$, the electric--electric correlators
(\ref{EE}) have a clear drop, while the magnetic--magnetic correlators
(\ref{BB}) stay almost unchanged, or show a slight increase.
\end{itemize}
In the next section we shall report on a quantitative analysis of the data
displayed in Figs. 1--8.

\newsection{Quantitative analysis}

\noindent
Inspired by our previous analyses of the correlators in the $T=0$ case
(see Refs. \cite{DiGiacomo92,npb97,plb97} and also \cite{EM99}),
we have performed best fits
to the lattice data for the correlators (\ref{corr-E}) and (\ref{corr-B})
at finite temperature $T$ (and at equal times, i.e., $u_4 = 0$),
with the functions (here $x = |\vec{u}|$):
\ba
{\cal D}^E(x) = A_0 {\rm e}^{-\mu_A x}
+ {a_0 \over x^4}  & ~~,~~ &
{\cal D}^E_1(x) = A_1 {\rm e}^{-\mu_A x}
+ {a_1 \over x^4}  ~,
\label{fit-E}
\\
{\cal D}^B(x) = B_0 {\rm e}^{-\mu_B x}
+ {b_0 \over x^4} & ~~,~~ &
{\cal D}^B_1(x) = B_1 {\rm e}^{-\mu_B x}
+ {b_1 \over x^4}  ~,
\label{fit-B}
\ea
where, of course, all the coefficients must be considered as functions of
the physical temperature $T$. The four independent functions (\ref{fit-E})
and (\ref{fit-B}) are written as the sum of a {\it non--perturbative}
exponential term plus a {\it perturbative--like} term behaving as $1/x^4$
(in fact, a term of this form is predicted by ordinary perturbation
theory).\footnote{Of course the coefficient of $1/x^4$ is
regularization--scheme dependent. In Eqs.~(\ref{fit-E}) and~(\ref{fit-B})
we refer to the lattice regularization scheme; other schemes could give
different values~\cite{Jamin98}.}
In Refs. \cite{npb97,plb97} the perturbative--like
term had the form $\sim e^{-\mu_a x} / x^4$: the exponential term
$e^{-\mu_a x}$ has been neglected here (i.e., we fix $\mu_a = 0$), since,
in the spirit of the Operator Product Expansion (OPE), we will concentrate
on the behaviour of the correlators at short distances.

Two quantities of physical interest enter in our best fits to the
lattice data:
\begin{itemize}
\item[(1)] The correlation length of the gluon field strengths,
defined as $\lambda_A = 1/{\mu_A}$ for the electric correlators and  
$\lambda_B = 1/{\mu_B}$ for the magnetic correlators.
({\it A priori}, we distinguish the two correlation lengths $\lambda_A$ and
$\lambda_B$ in the parametrization (\ref{fit-E})--(\ref{fit-B}).
However, as we shall be below, we have found that our data can be well
fitted using the same correlation length $\lambda_A = \lambda_B$ for
the electric and the magnetic correlators.)
\item[(2)] The gluon condensate, defined as
\be
G_2 \equiv \langle {\alpha_s \over \pi} :G^a_{\mu\nu} G^a_{\mu\nu}: \rangle
~~~~~~~~~ (\alpha_s = {g^2 \over 4\pi}) ~.
\label{G2}
\ee
\end{itemize}
These two quantities play an important role in phenomenology. 
The correlation length is relevant for the description of vacuum models
\cite{Dosch87,Dosch88,Simonov89}.

The gluon condensate was first introduced in Ref.~\cite{SVZ79}, 
in the context of the SVZ sum rules. As pointed out in Refs. \cite{plb97,EM99},
lattice provides us with a regularized determination of the correlators.
We shall briefly repeat here the argumentation originally reported in
Refs. \cite{plb97,EM99}. As in Ref.~\cite{SVZ79}, our correlators can be
given an OPE~\cite{Wilson69}:
\be
{1 \over 2\pi^2} {\cal D}_{\mu\nu,\mu\nu}(x) \mathop\sim_{x\to0}
C_{\bf 1}(x) \langle {\bf 1} \rangle + C_g(x) G_2 + 
\displaystyle\sum_{f=1}^{N_f} C_f(x) m_f \langle :\bar{q}_f q_f: \rangle
+ \ldots ~,
\label{OPE}
\ee
if we have $N_f$ quark flavours with masses $m_f$. (Of course, the last
term in Eq. (\ref{OPE}), i.e., the mixing to the quark condensates, is
absent in the {\it quenched} theory.)
The mixing to the identity operator $C_{\bf 1}(x)$ has a $1/x^4$
behaviour at small $x$. The mixings to the operators of dimension four 
$C_g(x)$ and $C_f(x)$ are expected to behave as constants for $x \to 0$.
Higher--order terms in the OPE (\ref{OPE}) are neglected.
The coefficients of the Wilson expansion are usually determined in 
perturbation theory and are known to be plagued by the so--called 
``infrared renormalons'' (see for example Ref. \cite{Mueller85} and 
references therein). 
In the same spirit of Ref.~\cite{SVZ79}, we shall disregard the
renormalon ambiguity, as we did in Ref.~\cite{plb97}.
With the normalization of Eq. (\ref{OPE}), this gives $C_g(0) \simeq 1$.
The contribution from the quark operators in (\ref{OPE})
can be neglected because it is higher order in $1/{\beta}$.
The gluon condensate is then, using the parametrization
(\ref{fit-E})--(\ref{fit-B}) for the correlators:
\be
G_2 = {3 \over \pi^2} (A_0 + A_1 + B_0 + B_1) ~.
\label{G2fit}
\ee
$G_2$ is the sum of an {\it electric} contribution,
that we shall call $G_2^{(ele)}$, plus a {\it magnetic} contribution,
that we shall call $G_2^{(mag)}$, which at non--zero temperature $T$
are in general different and should be distinguished:
\be
G_2^{(ele)} \equiv {g^2 \over \pi^2} \langle : \Tr[\vec{E}^2] : \rangle ~; ~~~
G_2^{(mag)} \equiv {g^2 \over \pi^2} \langle : \Tr[\vec{B}^2] : \rangle ~.
\label{G2em}
\ee
When using the parametrization (\ref{fit-E})--(\ref{fit-B}) for the
correlators, one easily finds that:
\be
G_2^{(ele)} = {3 \over \pi^2} (A_0 + A_1) ~; ~~~
G_2^{(mag)} = {3 \over \pi^2} (B_0 + B_1) ~.
\label{G2em-fit}
\ee
Let us discuss now the results obtained from the best fits to our data
with the functions (\ref{fit-E})--(\ref{fit-B}).
First of all, we have tried a best fit to the data for the magnetic
correlators (\ref{corr-B}) with the functions (\ref{fit-B}), where the
mass $\mu_B$ of the non--perturbative exponential terms has been put
equal to the corresponding value obtained in the $T=0$ case:
\ba
\mu_B &=& 4.53(7) ~{\rm fm}^{-1} ~~~ (quenched) ~; \nonumber \\
\mu_B &=& 3.5(2) ~{\rm fm}^{-1} ~~~ ({\rm full~QCD}) ~.
\label{mu-B}
\ea
The {\it quenched} value has been taken from Ref. \cite{npb97}, 
Eqs. (2.10) and (2.11); the correlators at $T = 0$ are reproduced
by the thick lines in Figs. 1--4.
Instead, the full--QCD value has been extracted from a best fit to the
new data at $T=0$ ($\beta = 5.55$ on a $16^4$ lattice), reported in Figs. 5--8.
The magnetic correlators are well fitted, up to distances of about
$x \simeq 0.8$ fm, by the functions (\ref{fit-B}) with the mass $\mu_B$
fixed to the value given in Eq. (\ref{mu-B}).

The results obtained for all the various cases are 
reported in Table I. From these results one sees that all the coefficients,
$B_0$, $B_1$, $b_0$ and $b_1$, are rather stable, when varying the temperature
$T$: only the non--perturbative coefficient $B_0$ shows a slight increase
when increasing $T$, so that, by virtue of Eq. (\ref{G2em-fit}), we can say
that the magnetic gluon condensate $G_2^{(mag)}$ slightly increases across
the transition at $T_c$ [see Fig. 9]. Let us also observe that the pure--gauge
magnetic gluon condensate is sensibly higher than the corresponding value in
the full--QCD case: this is in agreement with the fact that
the gluon condensate $G_2$ is expected to increase with the quark mass
\cite{NSVZ81}, tending towards the asymptotic (pure--gauge) value,
as already checked on data at $T = 0$~\cite{plb97}.
The stability of the perturbative coefficients $b_0$ and $b_1$ is expected
since they are UV--dominated cut--off terms.
More precisely, these perturbative coefficients are practically independent
on the lattice volume ($N_\sigma^3 \times N_\tau$), as one can verify by
direct computation \cite{CDM84}.
In general, (at a fixed lattice volume $N_\sigma^3 \times N_\tau$) they depend
on $\beta$ and it is known that $b_0 = {\cal O}(1/\beta^2)$ and
$b_1 = {\cal O}(1/\beta)$.
However, in the range of values of $T$ that we have considered,
$\Delta\beta/\beta \ll 1$ and the $\beta$ dependence of the perturbative
coefficients can be neglected within the errors.
For the same reasons, these magnetic perturbative
coefficients should be practically equal to the corresponding coefficients
$a_0$ and $a_1$ in the electric correlators (\ref{fit-E}).

Therefore, on the basis of these considerations, we have tried a best fit
to the data for the electric correlators (\ref{corr-E}) for distances from 3
up to 5--6 lattice spacings (corresponding approximately to the range of
physical distances $0.3 \div 0.6$ fm),\footnote{This is the range of
distances where we have data for the parallel electric correlator at all
temperatures.}
with the functions (\ref{fit-E}), where the perturbative coefficients
$a_0$ and $a_1$ have been fixed to the (weighted) average values of the
corresponding magnetic coefficients $b_0$ and $b_1$ reported in Table I
for all the temperatures that we have considered:
\ba
a_0 = 0.55(2) ~&;& ~~~ a_1 = 0.35(1) ~~~ (quenched) ~; \nonumber \\
a_0 = 0.64(2) ~&;& ~~~ a_1 = 0.33(1) ~~~ ({\rm full~QCD}) ~.
\label{a0a1}
\ea
Moreover, we have fixed the mass $\mu_A$ of the non--perturbative
exponential terms in (\ref{fit-E}) to the same value $\mu_B$ used for the
magnetic correlators [Eq. (\ref{mu-B})], which is in turn the value at $T=0$.
In Table II we report the results obtained for the perpendicular electric
correlator ${\cal D}^E_\perp$ [see Eq. (\ref{corr-E})].
The coefficient $A_0 + A_1$ of the non--perturbative part
of the correlator, which, by virtue of Eq. (\ref{G2em-fit}),
is proportional to the electric gluon condensate $G_2^{(ele)}$,
sharply decreases across the transition [see Fig. 10].
Again, we find that the pure--gauge electric gluon condensate is sensibly
higher than the corresponding value in the full--QCD case, in agreement with
the general claim done in Ref. \cite{NSVZ81}.

Finally, we have performed a best fit to the values for the difference
\be
{\cal D}_\perp^E (x) - {\cal D}_\parallel^E (x) =
-{x \over 2} {\partial{\cal D}_1^E \over \partial x} (x)
\label{diff-ele}
\ee
between the two quantities displayed in Figs. 2 and 1 respectively
(Figs. 6 and 5 for the full--QCD case). The results are reported in Table III.
Since (from previous experience at $T=0$) the quantity in the r.h.s. of Eq.
(\ref{diff-ele}) is expected to be dominated by the perturbative term,
in this case we have left also the coefficient $a_1$ as a free parameter
in the fit, in order to test the validity of the assumption that the
perturbative coefficients are temperature independent. The test is
perfect for the {\it quenched} case and reasonable for the full--QCD case.
One finds that the coefficient $A_1$ of the non--perturbative part
of the function ${\cal D}_1^E$ stays compatible with
zero, within the (large) errors, across the phase transition at
$T_c$. So the clear drop seen in the quantities ${\cal D}_\parallel^E$ and
${\cal D}_\perp^E$ across $T_c$ seems to be entirely due to the coefficient
$A_0$ of the non--perturbative part of the function ${\cal D}^E$ alone.
This result is consistent with Refs. \cite{npb97,DiGiacomo96} and again
confirms the conclusion of Refs. \cite{Simonov1,Simonov2,Simonov3},
where ${\cal D}^E$ (or, better, its non--perturbative part) was
related to the ({\it temporal}) string tension $\sigma_E$.
It was also shown in Refs. \cite{Simonov1,Simonov2} that ${\cal D}^B$
(or, better, its non--perturbative part) is
related to the {\it spatial} string tension $\sigma_s$.
Existing lattice results \cite{Bali93,Laermann95,Boyd96} indicate that
$\sigma_s$ is almost constant around $T_c$ and increases for $T \ge 2 T_c$:
this fact is in good agreement with the behaviour that we find for the
coefficient $B_0$ of the non--perturbative part of the function
${\cal D}^B$ (see Table I).

In summary, our best--fit analysis of the data leads to the following
conclusions:
\begin{itemize}
\item[(1)] The correlation lengths $\lambda_A = 1/{\mu_A}$ for the electric
correlators [Eq. (\ref{fit-E})] and  $\lambda_B = 1/{\mu_B}$ for the magnetic
correlators [Eq. (\ref{fit-B})] are equal and do not change across the
deconfining phase transition at $T_c$.
\item[(2)] The electric gluon condensate, defined in Eqs. (\ref{G2em}) and
(\ref{G2em-fit}), drops to zero at $T=T_c$, whilst the magnetic gluon
condensate is practically unchanged, showing a small increase
[see Figs. 9 and 10].
\item[(3)] As expected, the coefficients of the perturbative terms are
temperature independent and are the same for the magnetic and the electric
correlators.
\end{itemize}

\bigskip
\noindent {\bf Acknowledgements}
\smallskip

Part of this work was done using the CRAY T3E of the CINECA Inter University
Computing Centre (Bologna, Italy). We would like to thank the CINECA
for the kind and highly qualified technical assistance.

Stimulating discussions with Yuri Simonov are warmly acknowledged.

\vfill\eject

{\renewcommand{\Large}{\normalsize}
}

\vfill\eject

\noindent
\begin{center}
{\bf TABLE CAPTIONS}
\end{center}
\vskip 0.5 cm
\begin{itemize}
\item [\bf Tab.~I.] Results obtained from a best fit to the data of the
magnetic correlators (\ref{corr-B}) with the functions (\ref{fit-B}),
for the various temperatures that we have examined: ``q'' stands for
{\it ``quenched''} data, while ``f'' stands for ``full--QCD'' data.
The mass $\mu_B$ of the non--perturbative exponential terms has been put
equal to the corresponding value obtained in the $T=0$ case [Eq. (\ref{mu-B})].
An asterisk $(*)$ near the value of some parameter means that the parameter
was an input for the best--fit.
\bigskip
\item [\bf Tab.~II.] Results obtained from a best fit to the data of the
perpendicular electric correlator ${\cal D}^E_\perp$
[see Eq. (\ref{corr-E})] with the functions (\ref{fit-E}),
where the perturbative coefficients
$a_0$ and $a_1$ have been fixed to the (weighted) average values of the
corresponding magnetic coefficients $b_0$ and $b_1$ reported in Table I
for all the temperatures that we have considered [see Eq. (\ref{a0a1})].
Moreover, we have fixed the mass $\mu_A$ of the non--perturbative
exponential terms in (\ref{fit-E}) to the same value $\mu_B$ used for
the magnetic correlators [Eq. (\ref{mu-B})], i.e., to the corresponding
value obtained in the $T=0$ case. The notation used is the same as in Table I.
\bigskip
\item [\bf Tab.~III.] Results obtained from a best fit to the data of the
difference between the perpendicular electric correlator ${\cal D}^E_\perp$
and the parallel electric correlator ${\cal D}^E_\parallel$
[see Eqs. (\ref{diff-ele}) and (\ref{corr-E})] with the functions
(\ref{fit-E}), where the mass $\mu_A$ of the non--perturbative
exponential terms in (\ref{fit-E}) has been fixed to the same value
used in Table II. The notation used is the same as in Tables I and II.
\end{itemize}

\vfill\eject

\vskip 1cm

\centerline{\bf Table I}

\vskip 5mm

\moveleft 0.7 in
\vbox{\offinterlineskip
\halign{\strut
\vrule \hfil\quad $#$ \hfil \quad & 
\vrule \hfil\quad $#$ \hfil \quad & 
\vrule \hfil\quad $#$ \hfil \quad & 
\vrule \hfil\quad $#$ \hfil \quad & 
\vrule \hfil\quad $#$ \hfil \quad & 
\vrule \hfil\quad $#$ \hfil \quad & 
\vrule \hfil\quad $#$ \hfil \quad & 
\vrule \hfil\quad $#$ \hfil \quad \vrule \cr
\noalign{\hrule}
{\rm theory} &
T/T_c &
B_0 &
B_1 &
\mu_B &
b_0 &
b_1 &
\chi^2/N \cr
& & (10^8~{\rm MeV}^4) & (10^8~{\rm MeV}^4) & ({\rm fm}^{-1}) & & & \cr
\noalign{\hrule}
\noalign{\hrule}
{\rm q} & 0.952 & 1320(70) & 62(70) & 4.53~(*) & 0.54(5) & 0.28(3) & 0.7 \cr
\noalign{\hrule}
{\rm q} & 0.974 & 1290(73) & -50(70) & 4.53~(*) & 0.60(5) & 0.36(3) & 1.2 \cr
\noalign{\hrule}
{\rm q} & 1.007 & 1240(52) & -51(54) & 4.53~(*) & 0.57(4) & 0.36(3) & 0.6 \cr
\noalign{\hrule}
{\rm q} & 1.030 & 1436(63) & 24(77) & 4.53~(*) & 0.51(5) & 0.34(3) & 0.5 \cr
\noalign{\hrule}
{\rm q} & 1.065 & 1305(54) & -55(70) & 4.53~(*) & 0.56(4) & 0.37(3) & 1.1 \cr
\noalign{\hrule}
{\rm q} & 1.127 & 1455(49) & 91(67) & 4.53~(*) & 0.50(4) & 0.35(3) & 1.9 \cr
\noalign{\hrule}
{\rm q} & 1.261 & 1490(38) & 17(60) & 4.53~(*) & 0.54(3) & 0.36(2) & 0.5 \cr
\noalign{\hrule}
{\rm f} & 0.73  & 446(25)  &  37(16) & 3.5~(*) & 0.59(5) & 0.19(4) & 1.7 \cr
\noalign{\hrule}
{\rm f} & 0.94 & 461(23)  &  24(24) & 3.5~(*) & 0.66(5) & 0.30(3) & 1.4 \cr
\noalign{\hrule}
{\rm f} & 1.02 & 463(18) & -22(21) & 3.5~(*) & 0.63(4) & 0.37(2) & 0.63 \cr
\noalign{\hrule}
{\rm f} & 1.18 & 510(30) & -11(27) & 3.5~(*) & 0.64(3) & 0.40(3) & 0.23 \cr
\noalign{\hrule}
{\rm f} & 1.48 & 574(35) & 8(34) & 3.5~(*) & 0.65(3) & 0.38(2) & 1.6 \cr
\noalign{\hrule}
}}

\vskip 1cm

\centerline{\bf Table II}

\vskip 5mm

\moveright .2 in
\vbox{\offinterlineskip
\halign{\strut
\vrule \hfil\quad $#$ \hfil \quad & 
\vrule \hfil\quad $#$ \hfil \quad & 
\vrule \hfil\quad $#$ \hfil \quad & 
\vrule \hfil\quad $#$ \hfil \quad & 
\vrule \hfil\quad $#$ \hfil \quad & 
\vrule \hfil\quad $#$ \hfil \quad \vrule \cr
\noalign{\hrule}
{\rm theory} &
T/T_c &
A_0 + A_1 &
\mu_A &
a_0 + a_1 &
\chi^2/N \cr
& & (10^8~{\rm MeV}^4) & ({\rm fm}^{-1}) & & \cr
\noalign{\hrule}
\noalign{\hrule}
{\rm q} & 0.952 & 1193(30) & 4.53~(*) & 0.90~(*) & 0.2 \cr
\noalign{\hrule}
{\rm q} & 0.974 & 930(25) & 4.53~(*) & 0.90~(*) & 1.6 \cr
\noalign{\hrule}
{\rm q} & 1.007 & 596(31) & 4.53~(*) & 0.90~(*) & 2.1 \cr
\noalign{\hrule}
{\rm q} & 1.030 & 318(42) & 4.53~(*) & 0.90~(*) & 0.9 \cr
\noalign{\hrule}
{\rm q} & 1.065 & 197(27) & 4.53~(*) & 0.90~(*) & 0.2 \cr
\noalign{\hrule}
{\rm q} & 1.127 & 56(60) & 4.53~(*) & 0.90~(*) & 0.3 \cr
\noalign{\hrule}
{\rm q} & 1.261 & -88(57) & 4.53~(*) & 0.90~(*) & 0.4 \cr
\noalign{\hrule}
{\rm f} & 0.73 & 381(10) & 3.5~(*) & 0.97~(*) & 0.2 \cr
\noalign{\hrule}
{\rm f} & 0.94 & 413(20) & 3.5~(*) &  0.97~(*) & 0.84 \cr
\noalign{\hrule}
{\rm f} & 1.02 & 288(25) & 3.5~(*) &  0.97~(*) & 1.9 \cr
\noalign{\hrule}
{\rm f} & 1.18 & 186(30) & 3.5~(*) &  0.97~(*) & 3.4 \cr
\noalign{\hrule}
{\rm f} & 1.48 & 43(25) & 3.5~(*) &  0.97~(*) & 3.3 \cr
\noalign{\hrule}
}}

\vfill\eject

\vskip 1cm

\centerline{\bf Table III}

\vskip 5mm

\vbox{\offinterlineskip
\halign{\strut
\vrule \hfil\quad $#$ \hfil \quad & 
\vrule \hfil\quad $#$ \hfil \quad & 
\vrule \hfil\quad $#$ \hfil \quad & 
\vrule \hfil\quad $#$ \hfil \quad & 
\vrule \hfil\quad $#$ \hfil \quad & 
\vrule \hfil\quad $#$ \hfil \quad \vrule \cr
\noalign{\hrule}
{\rm theory} &
T/T_c &
A_1 &
\mu_A &
a_1 &
\chi^2/N \cr
& & (10^8~{\rm MeV}^4) & ({\rm fm}^{-1}) & & \cr
\noalign{\hrule}
\noalign{\hrule}
{\rm q} & 0.952 & -70(305) & 4.53~(*) & 0.34(9) & 3 \times 10^{-3} \cr
\noalign{\hrule}
{\rm q} & 0.974 & 243(285) & 4.53~(*) & 0.34(8) & 1 \times 10^{-3} \cr
\noalign{\hrule}
{\rm q} & 1.007 & 508(332) & 4.53~(*) & 0.36(7) & 2 \times 10^{-2} \cr
\noalign{\hrule}
{\rm q} & 1.030 & 570(587) & 4.53~(*) & 0.37(15) & 1.4 \times 10^{-2} \cr
\noalign{\hrule}
{\rm q} & 1.065 & 544(790) & 4.53~(*) & 0.38(15) & 6 \times 10^{-2} \cr
\noalign{\hrule}
{\rm q} & 1.127 & 502(829) & 4.53~(*) & 0.39(15) & 1 \times 10^{-1} \cr
\noalign{\hrule}
{\rm q} & 1.261 & 23(1733) & 4.53~(*) & 0.46(25) & 2 \times 10^{-2} \cr
\noalign{\hrule}
{\rm f} & 0.73 & 29(61) & 3.5~(*) & 0.20(7) & N = 0 \cr
\noalign{\hrule}
{\rm f} & 0.94 & 5(70) & 3.5~(*) & 0.33(7) & 0.01 \cr
\noalign{\hrule}
{\rm f} & 1.02 & 153(80) & 3.5~(*) & 0.37(7) & 0.1  \cr
\noalign{\hrule}
{\rm f} & 1.18 & -10(40) & 3.5~(*) & 0.53(6) & 1.9 \cr
\noalign{\hrule}
{\rm f} & 1.48 & -110(80) & 3.5~(*) & 0.55(6) & 1.2 \cr
\noalign{\hrule}
}}

\newpage

\noindent
\begin{center}
{\bf FIGURE CAPTIONS}
\end{center}
\vskip 0.5 cm
\begin{itemize}
\item [\bf Fig.~1.] The quantity ${\cal D}_\parallel^E$ [Eq. (\ref{corr-E})],
in units of MeV$^4$, versus the physical distance (in fm),
for different values of $T/T_c$ in the {\it quenched} theory.
The thick continuum line has been obtained using the parameters of the best fit
to the data at $T=0$ [Ref. \cite{npb97}, Eqs. (2.10) and (2.11)].
\bigskip
\item [\bf Fig.~2.] The quantity ${\cal D}_\perp^E$ [Eq. (\ref{corr-E})],
in units of MeV$^4$, versus the physical distance (in fm),
for different values of $T/T_c$ in the {\it quenched} theory.
The thick continuum line has been obtained using the parameters of the best fit
to the data at $T=0$ [Ref. \cite{npb97}, Eqs. (2.10) and (2.11)].
\bigskip
\item [\bf Fig.~3.] The quantity ${\cal D}_\parallel^B$ [Eq. (\ref{corr-B})],
in units of MeV$^4$, versus the physical distance (in fm),
for different values of $T/T_c$ in the {\it quenched} theory.
The thick continuum line has been obtained using the parameters of the best fit
to the data at $T=0$ [Ref. \cite{npb97}, Eqs. (2.10) and (2.11)].
\bigskip
\item [\bf Fig.~4.] The quantity ${\cal D}_\perp^B$ [Eq. (\ref{corr-B})],
in units of MeV$^4$, versus the physical distance (in fm),
for different values of $T/T_c$ in the {\it quenched} theory.
The thick continuum line has been obtained using the parameters of the best fit
to the data at $T=0$ [Ref. \cite{npb97}, Eqs. (2.10) and (2.11)].
\bigskip
\item [\bf Fig.~5.] The quantity ${\cal D}_\parallel^E$ [Eq. (\ref{corr-E})],
in units of MeV$^4$, versus the physical distance (in fm),
for different values of $T/T_c$ and for $T = 0$ in full QCD.
\bigskip
\item [\bf Fig.~6.] The quantity ${\cal D}_\perp^E$ [Eq. (\ref{corr-E})],
in units of MeV$^4$, versus the physical distance (in fm),
for different values of $T/T_c$ and for $T = 0$ in full QCD.
\bigskip
\item [\bf Fig.~7.] The quantity ${\cal D}_\parallel^B$ [Eq. (\ref{corr-B})],
in units of MeV$^4$, versus the physical distance (in fm),
for different values of $T/T_c$ and for $T = 0$ in full QCD.
\bigskip
\item [\bf Fig.~8.] The quantity ${\cal D}_\perp^B$ [Eq. (\ref{corr-B})],
in units of MeV$^4$, versus the physical distance (in fm),
for different values of $T/T_c$ and for $T = 0$ in full QCD.
\bigskip
\item [\bf Fig.~9.] The magnetic gluon condensate $G_2^{(mag)}(T)$ [see Eqs.
(\ref{G2em}) and (\ref{G2em-fit})], in units of $G_2^{(mag)}(T=0)$, versus
$T/T_c$. The black circles refer to the {\it quenched} case, while the
white circles refer to the full--QCD case.
\bigskip
\item [\bf Fig.~10.] The electric gluon condensate $G_2^{(ele)}(T)$ [see Eqs.
(\ref{G2em}) and (\ref{G2em-fit})], in units of $G_2^{(ele)}(T=0)$, versus
$T/T_c$. The notation is the same as in Fig. 9.
\end{itemize}

\vfill\eject

\pagestyle{empty}

\centerline{\bf Figure 1}
\vskip 4truecm
\begin{figure}[htb]
\vskip 4.5truecm
\includegraphics{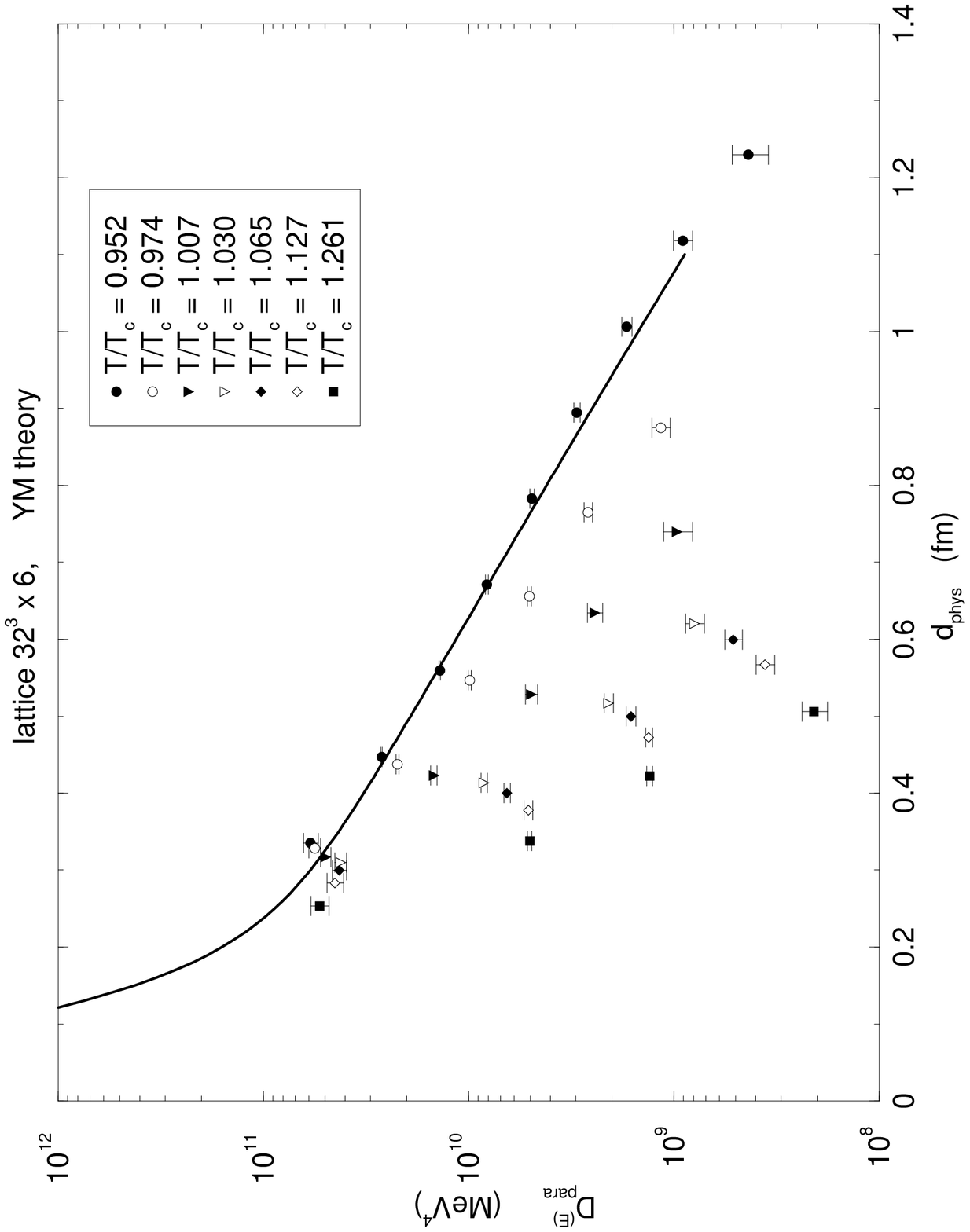}
\end{figure}

\vfill\eject

\centerline{\bf Figure 2}
\vskip 4truecm
\begin{figure}[htb]
\vskip 4.5truecm
\includegraphics{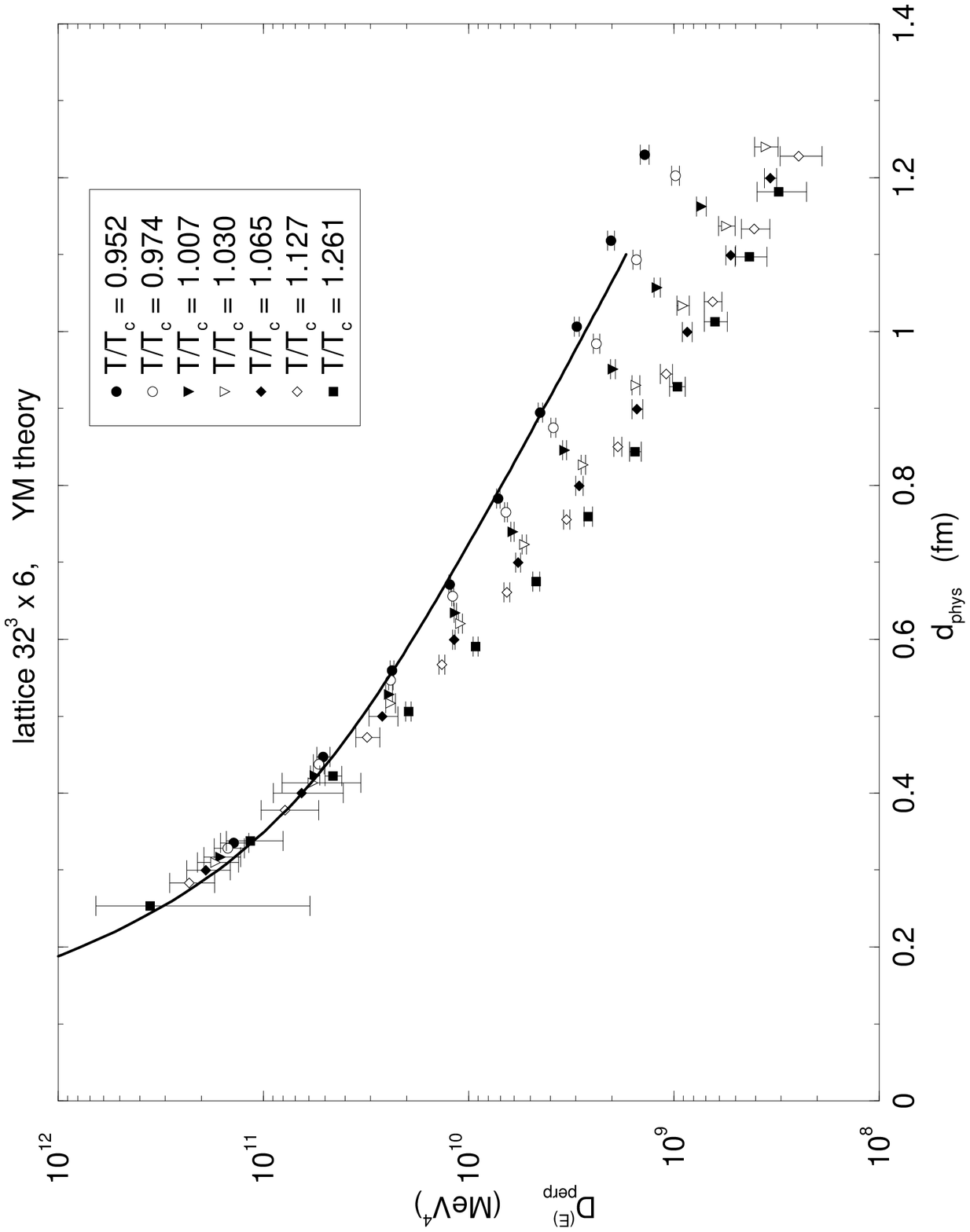}
\end{figure}

\vfill\eject

\centerline{\bf Figure 3}
\vskip 4truecm
\begin{figure}[htb]
\vskip 4.5truecm
\includegraphics{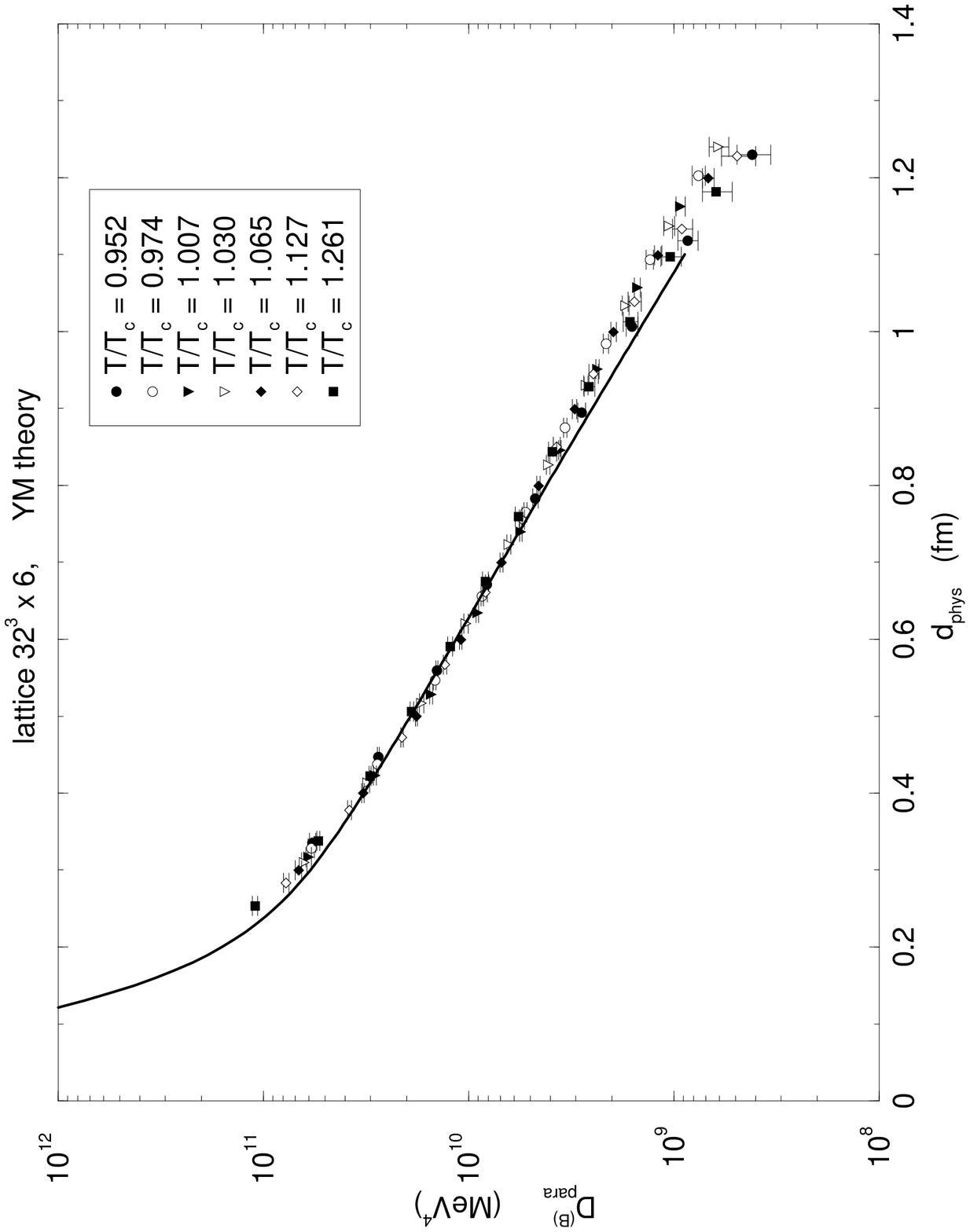}
\end{figure}

\vfill\eject

\centerline{\bf Figure 4}
\vskip 4truecm
\begin{figure}[htb]
\vskip 4.5truecm
\includegraphics{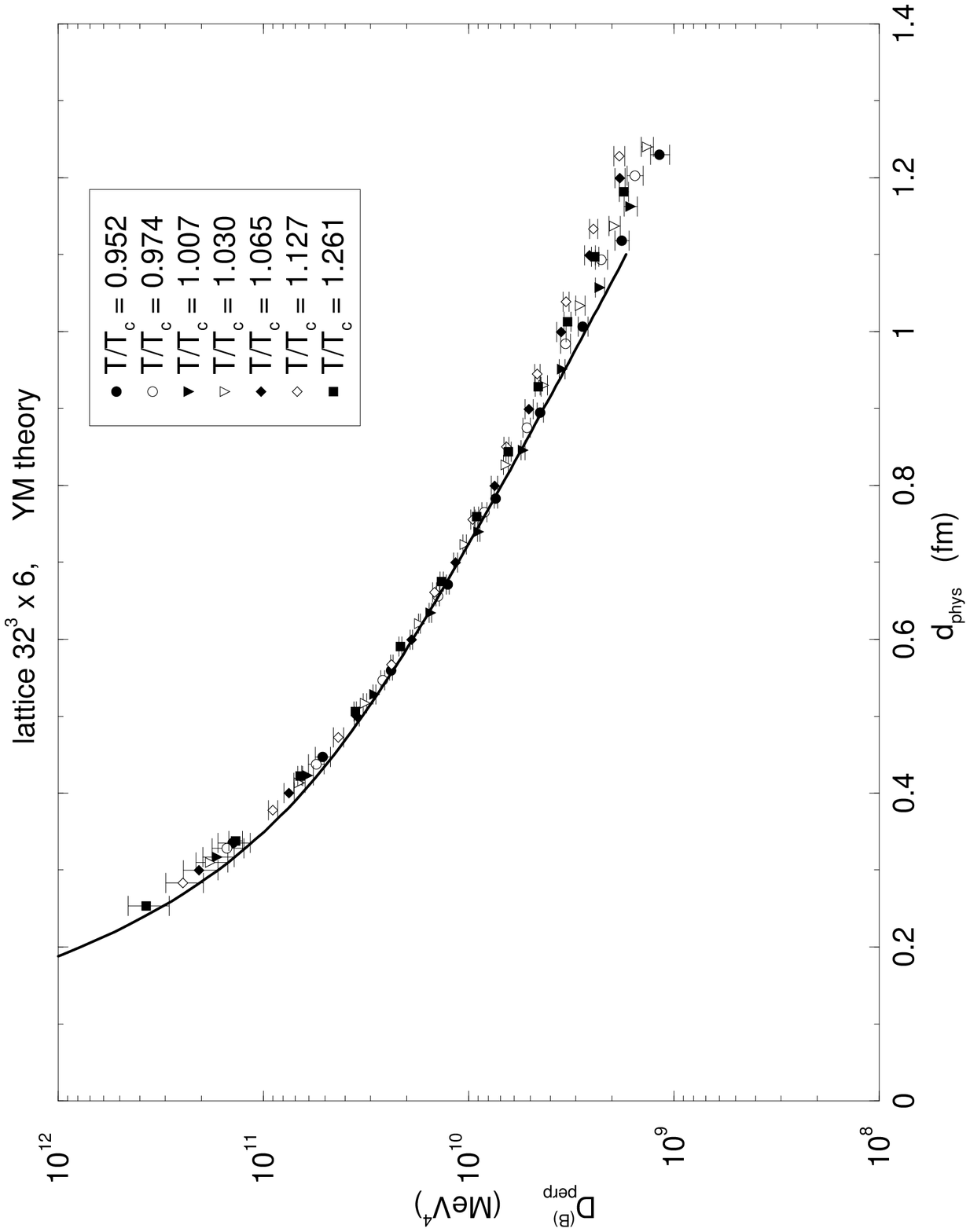}
\end{figure}

\vfill\eject

\centerline{\bf Figure 5}
\vskip 4truecm
\begin{figure}[htb]
\vskip 4.5truecm
\includegraphics{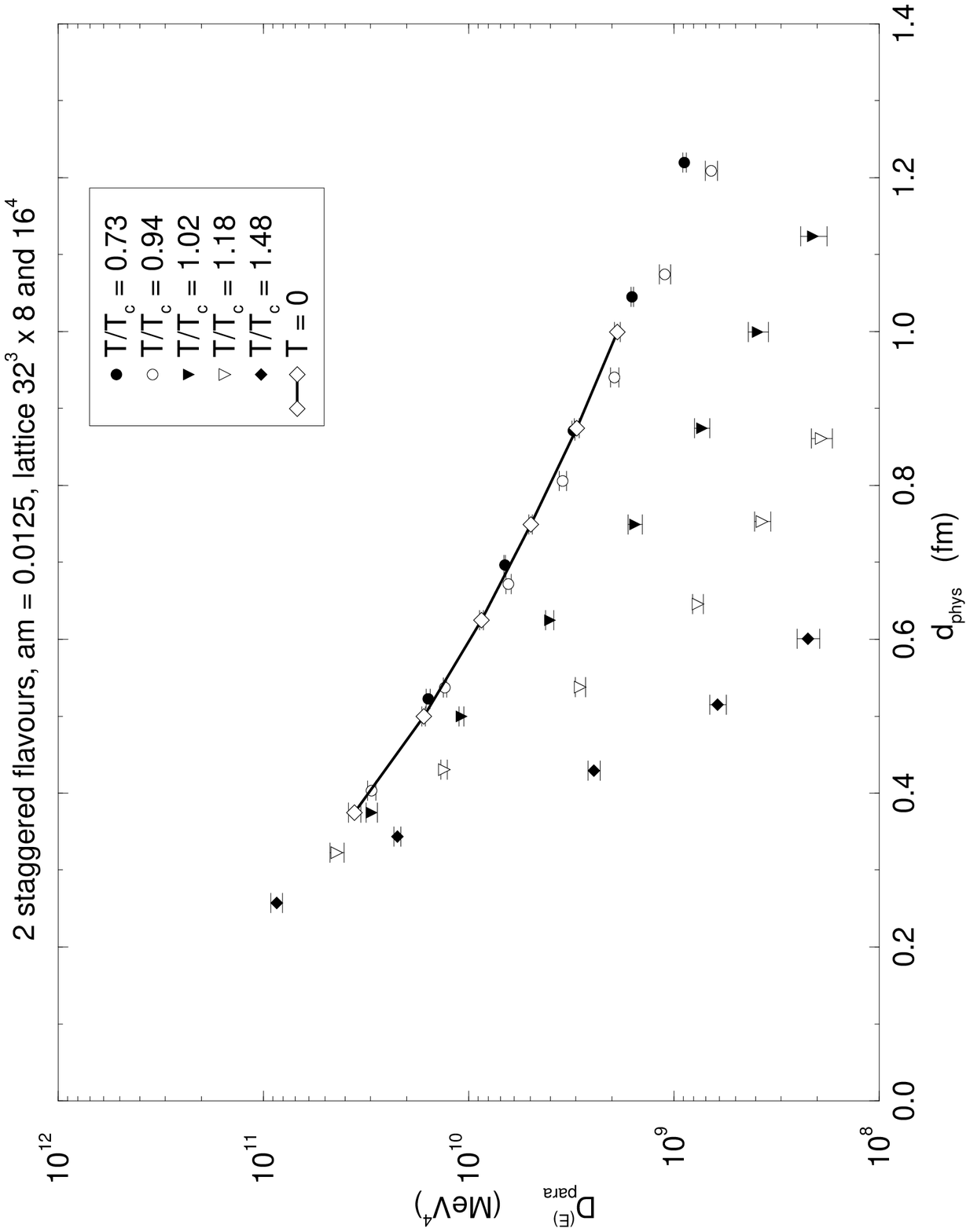}
\end{figure}

\vfill\eject

\centerline{\bf Figure 6}
\vskip 4truecm
\begin{figure}[htb]
\vskip 4.5truecm
\includegraphics{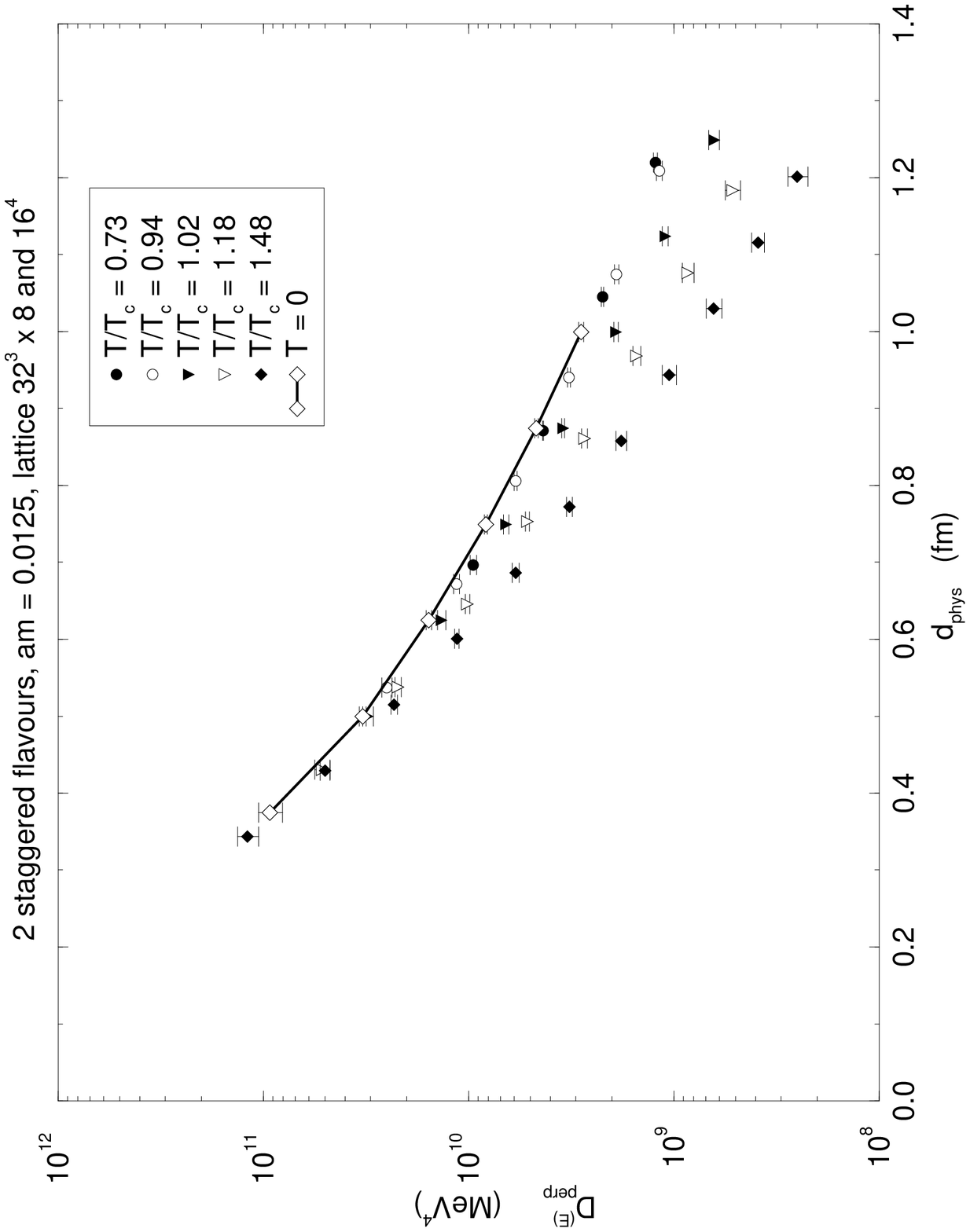}
\end{figure}

\vfill\eject

\centerline{\bf Figure 7}
\vskip 4truecm
\begin{figure}[htb]
\vskip 4.5truecm
\includegraphics{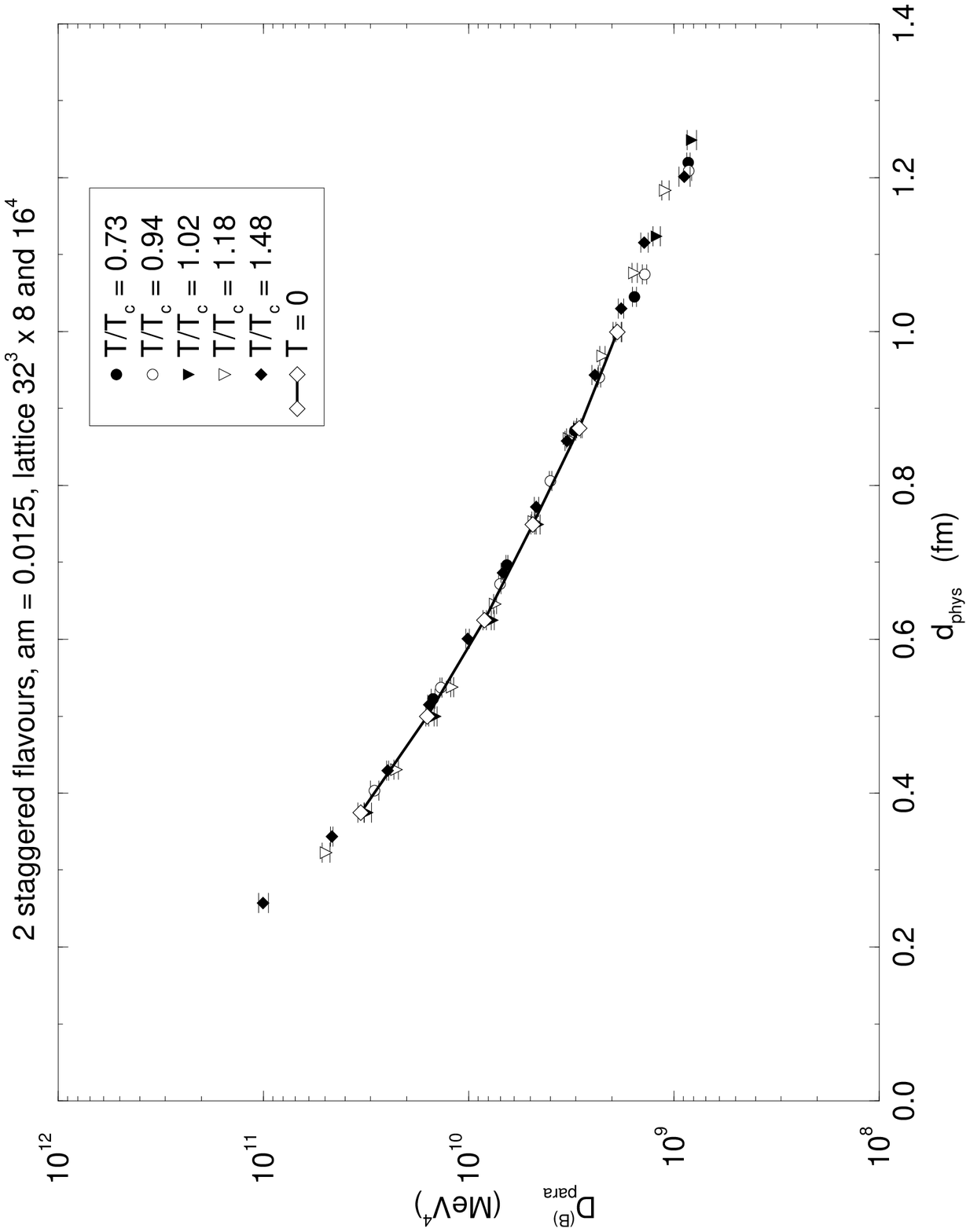}
\end{figure}

\vfill\eject

\centerline{\bf Figure 8}
\vskip 4truecm
\begin{figure}[htb]
\vskip 4.5truecm
\includegraphics{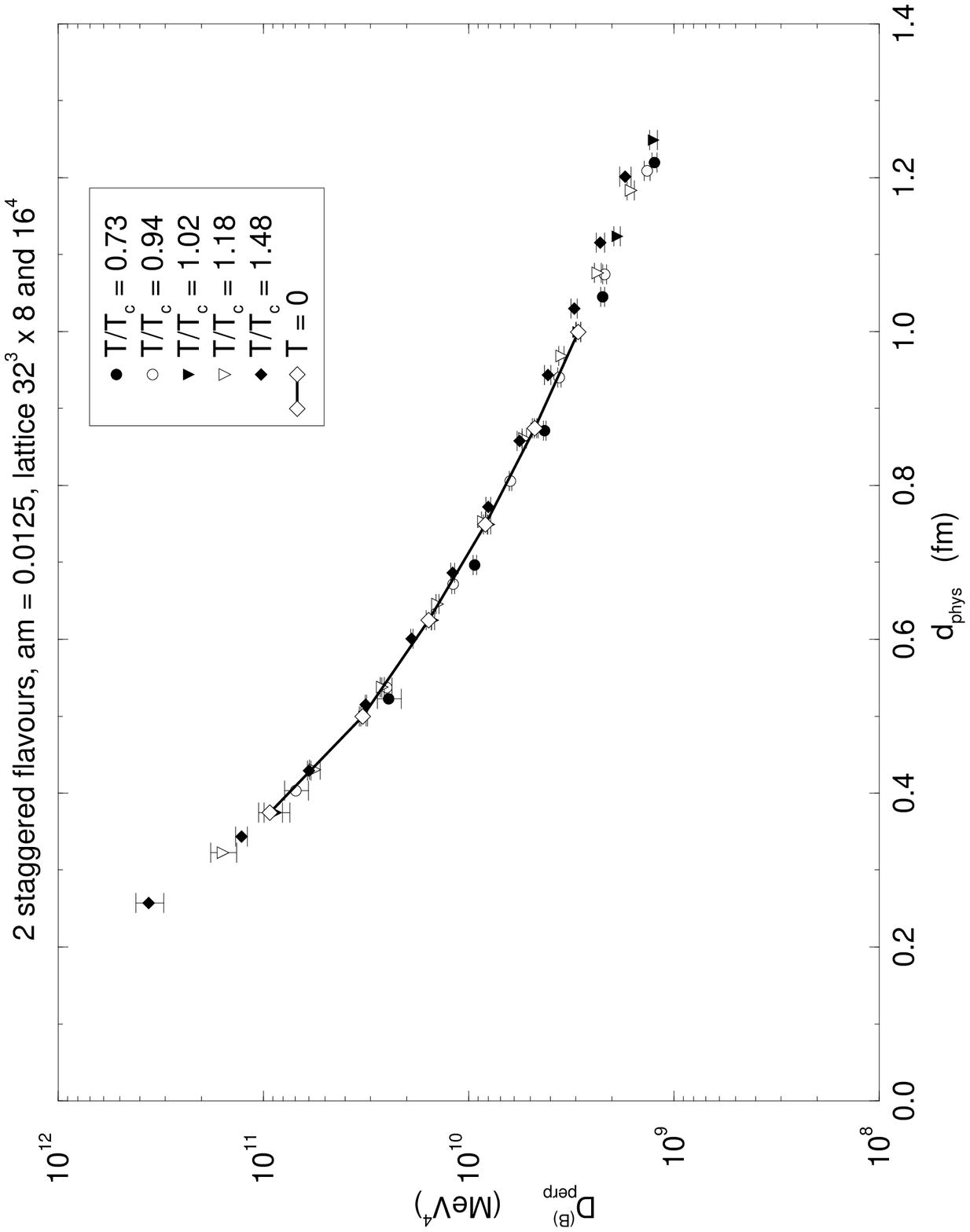}
\end{figure}

\vfill\eject

\centerline{\bf Figure 9}
\vskip 4truecm
\begin{figure}[htb]
\vskip 4.5truecm
\includegraphics{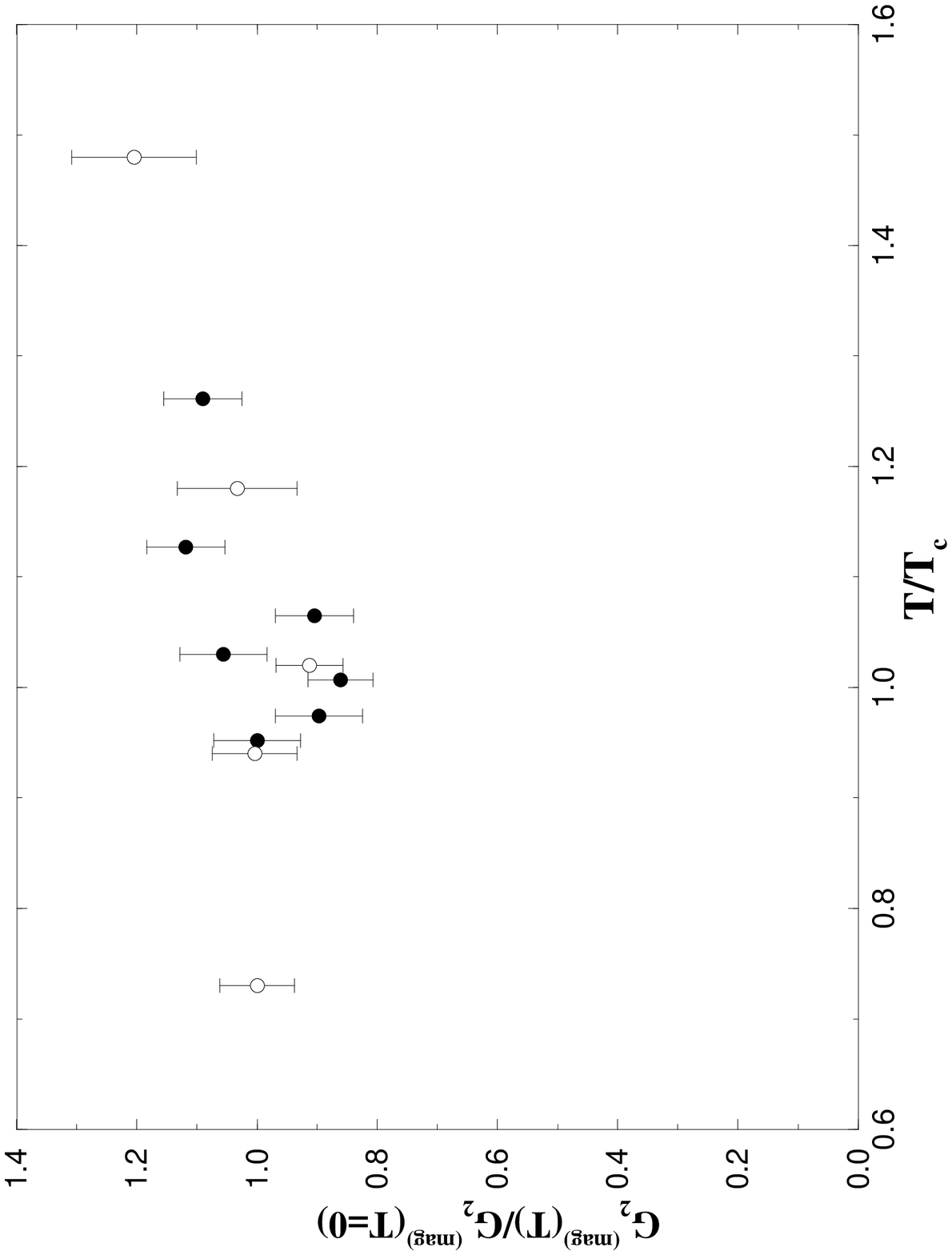}
\end{figure}

\vfill\eject

\centerline{\bf Figure 10}
\vskip 4truecm
\begin{figure}[htb]
\vskip 4.5truecm
\includegraphics{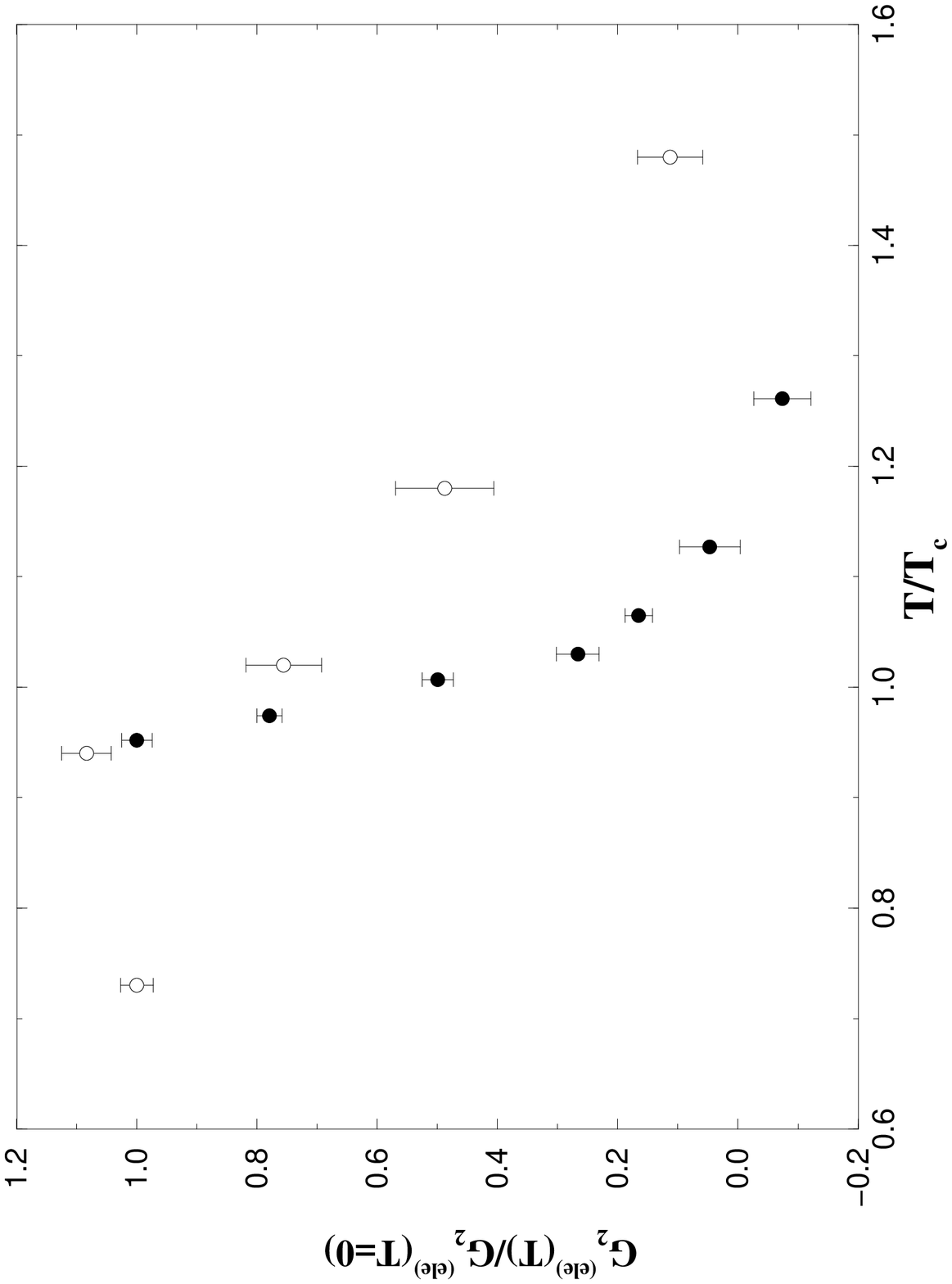}
\end{figure}

\vfill\eject


\begin{thebibliography}{99}
\bibitem{Dosch87}
H.G. Dosch, Phys. Lett. B {\bf 190} (1987) 177.
\bibitem{Dosch88}
H.G. Dosch and Yu.A. Simonov, Phys. Lett. B {\bf 205} (1988) 339.
\bibitem{Simonov89}
Yu.A. Simonov, Nucl. Phys. B {\bf 324} (1989) 67.
\bibitem{Simonov1}
Yu.A. Simonov, JETP Lett. {\bf 54} (1991) 249.
\bibitem{Simonov2}
Yu.A. Simonov, JETP Lett. {\bf 55} (1992) 627;
Yad. Fiz. {\bf 58} (1995) 357.
\bibitem{Simonov3}
Yu.A. Simonov and E.L. Gubankova, Phys. Lett. B {\bf 360} (1995) 93.
\bibitem{DDSS02}
A. Di Giacomo, H.G. Dosch, V.I. Shevchenko and Yu.A. Simonov,
Phys. Rep. C {\bf 372} (2002) 319.
\bibitem{npb97}
A. Di Giacomo, E. Meggiolaro and H. Panagopoulos, Nucl. Phys. B {\bf 483}
(1997) 371.
\bibitem{DiGiacomo96}
A. Di Giacomo, E. Meggiolaro and H. Panagopoulos, Pisa preprint IFUP--TH
14/96 (1996); Cyprus preprint UCY--PHY--96/6 (1996); hep--lat/9603018.
\bibitem{Simonov4}
D.S. Kuzmenko and Yu.A. Simonov, Phys. Atom. Nucl. {\bf 64} (2001) 1887;
Yad. Fiz. {\bf 64} (2001) 1971.
\bibitem{Michael88}
C. Michael and M. Teper, Phys. Lett. B {\bf 206} (1988) 299.
\bibitem{Bali-Schilling93}
G.S. Bali and K. Schilling, Phys. Rev. D {\bf 47} (1993) 661.
\bibitem{plb2002}
A. Di Giacomo and E. Meggiolaro, Phys. Lett. B {\bf 537} (2002) 173.
\bibitem{pisa2000}
B. Alles, M. D'Elia and A. Di Giacomo, Phys. Lett. B {\bf 483} (2000) 189.  
\bibitem{DiGiacomo92}
A. Di Giacomo and H. Panagopoulos, Phys. Lett. B {\bf 285} (1992) 133.
\bibitem{Boyd96}
G. Boyd, J. Engels, F. Karsch, E. Laermann, C. Legeland, M. L\"utgemeier and
B. Petersson, Nucl. Phys. B {\bf 469} (1996) 419.
\bibitem{gott93}
S. Gottlieb, A. Krasnitz, U.M. Heller, A.D. Kennedy, J.B. Kogut,
R.L. Renken, D.K. Sinclair, R.L. Sugar, D. Toussaint and K.C. Wang,
Phys. Rev. D {\bf 47} (1993) 3619.
\bibitem{gott87}
S. Gottlieb, W. Liu, D. Toussaint, R.L. Renken and R.L. Sugar,
Phys. Rev. D {\bf 35} (1987) 2531.
\bibitem{plb97}
M. D'Elia, A. Di Giacomo and E. Meggiolaro, Phys. Lett. B {\bf 408} (1997) 315.
\bibitem{EM99}
E. Meggiolaro, Phys. Lett. B {\bf 451} (1999) 414.
\bibitem{Jamin98}
M. Eidem\"uller and M. Jamin, Phys. Lett. B {\bf 416} (1998) 415.
\bibitem{SVZ79}
M.A. Shifman, A.I. Vainshtein, and V.I. Zakharov, Nucl. Phys. B {\bf 147}
(1979) 385; 448; 519.
\bibitem{Wilson69}
K.G. Wilson, Phys. Rev. {\bf 179} (1969) 1499.
\bibitem{Mueller85}
A.H. Mueller, Nucl. Phys. B {\bf 250} (1985) 327;\\
A.H. Mueller, in {\it QCD, 20 Years Later}, edited by P.M. Zerwas and
H.A. Kastrup (World Scientific, Singapore, 1993).
\bibitem{NSVZ81}
V.A. Novikov, M.A. Shifman, A.I. Vainshtein, and V.I. Zakharov, 
Nucl. Phys. B {\bf 191} (1981) 301.
\bibitem{CDM84}
M. Campostrini, A. Di Giacomo and G. Mussardo, Z. Phys. C {\bf 25} (1984) 173.
\bibitem{Bali93}
G.S. Bali, J. Fingberg, U.M. Heller, F. Karsch and K. Schilling,
Phys. Rev. Lett. {\bf 71} (1993) 3059.
\bibitem{Laermann95}
E. Laermann, Nucl. Phys. B (Proc. Suppl.) {\bf 42} (1995) 120.
\end{thebibliography}
\end{document}